\documentclass[prd,aps,showpacs,tightenlines,superscriptaddress,amsmath,amssymb,amsfonts,twocolumn]{revtex4-2}
\usepackage{amssymb,amsmath,amsthm,amsbsy,epsfig,color,graphicx,times,physics}
\usepackage[ansinew]{inputenc}
\usepackage[english]{babel}
\usepackage{color}
\usepackage{braket}
\usepackage{tabularx}
\usepackage{array}
\usepackage{hyperref}
\usepackage{slashed}
\usepackage{enumitem}
\usepackage{float}

\begin{document}
	\title{Probing Mirror Neutrons and Dark Matter through Cold Neutron Interferometry}
	\author{Antonio Capolupo}
	\email{capolupo@sa.infn.it}
	\affiliation{Dipartimento di Fisica ``E.R. Caianiello'' Universit\`{a} di Salerno, and INFN -- Gruppo Collegato di Salerno, Via Giovanni Paolo II, 132, 84084 Fisciano (SA), Italy}
	
	\author{Gabriele Pisacane}
	\email{gpisacane@unisa.it}
	\affiliation{Dipartimento di Fisica ``E.R. Caianiello'' Universit\`{a} di Salerno, and INFN -- Gruppo Collegato di Salerno, Via Giovanni Paolo II, 132, 84084 Fisciano (SA), Italy}
	
	\author{Aniello Quaranta}
	\email{anquaranta@unisa.it}
	\affiliation{Dipartimento di Fisica ``E.R. Caianiello'' Universit\`{a} di Salerno, and INFN -- Gruppo Collegato di Salerno, Via Giovanni Paolo II, 132, 84084 Fisciano (SA), Italy}

	\author{Francesco Romeo}
	\email{fromeo@unisa.it}
	\affiliation{Dipartimento di Fisica ``E.R. Caianiello'' Universit\`{a} di Salerno, and INFN -- Gruppo Collegato di Salerno, Via Giovanni Paolo II, 132, 84084 Fisciano (SA), Italy}

    \begin{abstract}

     We propose a novel neutron interferometry setup to explore the potential existence of mirror neutrons, a candidate for dark matter. Our work demonstrates that if mirror neutrons exist, neutrons will acquire an observable geometric phase due to mixing with these mirror counterparts. This geometric phase, detectable through our interferometric setup, could serve as a direct probe for the presence of mirror matter particles. Additionally, this investigation could shed light on unresolved issues in particle physics, such as the neutron lifetime puzzle. We discuss the setup's versatility and limitations, showing its capability to explore a wide range of parameters in neutron interferometry and potentially uncover new physics.

    \end{abstract}

    \maketitle

	Physics beyond the Standard Model is called for by a substantial amount of experimental evidence, ranging from neutrino oscillations \cite{Neut1,Neut2,Neut3,Neut4,Neut5,Neut6,Neut7,Neut8,Neut9,Neut10}, to particle physics anomalies such as the muon $g-2$ \cite{Muon1,Muon2} and the neutron lifetime puzzle  \cite{Czarnecki,Mumm,Berezhiani2019}, to dark matter \cite{Rubin1,Rubin2,Trimble1987,Corbelli2000,Clesse2018,Capolupo2021,Salucci2021,Capolupo2020}. The physics underlying one of the above phenomena often offers an explanation also for some of the others. It is the case for instance of axions, simultaneously solving the strong CP problem \cite{Peccei1,Peccei2,Weinberg,Wilczek,Raffelt,CapolupoAx2020,Marsh} and representing an ideal dark matter candidate. Many extensions of the standard model, including supersymmetry and mirror matter theory \cite{YangLee1956,Kobzarev1966,Blinnikov1982,Berezhiani2004,Hodges1993,Foot2014,Hao2022,Hostert2023,Berezhiani2006}, yield novel particles that may be responsible for the missing matter in the universe. Mirror matter theory, besides the appealing restoration of the parity simmetry at high energy, may account for the observed neutron lifetime anomaly \cite{Berezhiani2019}. Remarkably mirror matter may reveal itself in the shape of neutron-mirror neutron mixing \cite{Berezhiani2019,Hostert2023,Berezhiani2006}, which we aim to probe in this work.

On the other hand, neutron interferometry \cite{RauchBook,Werner1979,Wagh1990,Allman1997,Filipp2005,CapolupoAx2021} has proven to be an exceptional tool for examining particle interactions and for analyzing their quantum features.  Neutron interferometry  has also allowed  to validate numerous theoretical predictions, as for example the existence of the  Berry phase \cite{Berry,Berry1} and of the non cyclic geometric phase \cite{Mukunda}.

Here we show that a geometric phase \cite{Geo1,Geo2,Geo3,Geo4,Mukunda,Capolupo2018} for the neutron naturally emerges due to the mixing with mirror neutron, and we propose a cold neutron interferometry setup to reveal it. We discuss how the interferometer may probe a significant portion of parameter space, possibly putting stringent bounds on the mirror matter model and eventually uncovering this hidden gauge sector.
	
	The idea that there may be a hidden gauge sector was first proposed in \cite{YangLee1956} to restore parity. Among the possible hidden sectors, mirror matter has been extensively discussed \cite{Kobzarev1966,Blinnikov1982,Berezhiani2004,Hodges1993,Foot2014,Hostert2023,Hao2022,Berezhiani2006,Berezhiani2019} and has structurally the same content as ordinary matter. Ordinary and mirror particles interact gravitationally and possibly through weak interactions, but they are entirely independent pertaining strong and electromagnetic interactions.
	They constitute parallel sectors, each governed by the same gauge symmetry group $G$ (where in the simplest case, $G = G_{SM} \equiv SU(3)_c \otimes SU(2)_L \otimes U(1)_Y$). This implies that the comprehensive gauge group is $G \otimes G$. Alongside the standard Higgs doublet $\phi$ there is a corresponding mirror Higgs doublet $\phi'$. If the mirror symmetry is spontaneously broken \cite{Lee1974} the two Higgs fields acquire different vacuum expectation values, $\langle \phi \rangle \neq \langle \phi' \rangle$, resulting in distinct masses for ordinary and mirror particles.
	In this framework, a universal mixing process between ordinary and mirror neutral hadrons can take place, enabling matter-mirror oscillations as a result of spontaneous breaking of mirror symmetry. For neutral hadrons such as neutrons, the physical states (neutron $n$ and mirror neutron $n'$) are not aligned with the definite mass states $n_1$ and $n_2$. As a consequence the $n-n'$ oscillation can be described using an approach akin to the one employed in modeling ordinary two-flavour neutrino oscillations \cite{Berezhiani2006}.\\
	The physical states $n$ and $n'$ are related to the mass states through a rotation: $\begin{pmatrix}
		n \\
		 {n'}
	\end{pmatrix}
	=
    \begin{pmatrix}
     \cos \theta & \sin \theta \\
     - \sin \theta & \cos \theta \\
    \end{pmatrix}
	\begin{pmatrix}
		n_1 \\
		n_2
	\end{pmatrix}$, where $\theta $ is the mixing angle. The mass term of the Hamiltonian in the $n,n'$ basis, for a given spin polarization $s$ reads
\begin{equation}\label{H}
H_s=	\begin{pmatrix}
		m_n+\Delta E_s & \epsilon_{n n'} \\
		\epsilon_{n n'} & m_n+\delta m
	\end{pmatrix}
\end{equation}
    where $m_n$ is the neutron mass, $\epsilon_{n n'}$ is the mixing amplitude and $\delta m = m_{n'} - m_n$ is the in-vacuum $n-n'$ mass splitting. $\Delta E_s$ is the energy shift due to neutron interactions with external fields. The only term relevant to our discussion is the dipole coupling to an external magnetic field $\pmb{B}$, which, assuming spin polarized along $\hat{B}$, is $\Delta E_s = -s\mu_{n}B$. Mixing is enhanced for spin aligned with the magnetic field $s=1$. We shall focus on this case, dropping the spin index and denoting $\Delta E = -\mu_{n}B= |\mu_{n}B|$.
    It should be emphasized that the model parameters $\delta m $ and $\epsilon_{n n'}$ have not yet been measured experimentally, and only upper bounds have been placed on them $\delta m \lesssim 10^{-7} \mathrm{eV}, \epsilon_{nn'} \lesssim 10^{-9} \mathrm{eV}$ \cite{Berezhiani2019}.\\
    The Hamiltonian \eqref{H} is readily diagonalized with mixing angle $\tan\left( 2 \theta \right) =\frac{2 \epsilon_{n n'}}{\delta m - \Delta E}$ and eigenvalues $m_{1,2} =\frac{1}{2} \left( 2 m_n + \Delta E + \delta m \mp \sqrt{\left( \Delta E - \delta m \right)^2 + 4 \epsilon^2_{n n'}}  \right)$. Considered that the relativistic energy for a free particle of mass $m_j$ is $\omega_j=\sqrt{m_j ^2 + k^2}$ with $j=1,2$, $k=|\pmb{k}|$ and $\pmb{k}$ the particle 3-momentum, the time evolution of the mass states is trivially $\ket{n_j (t)} = e^{- i \omega_j t} \ket{n_j}$, where it is understood that $\ket{n_j} = \ket{n_j (t=0)}$. Dropping an irrelevant common phase factor, the physical states are simply the orthogonal combinations
	 \begin{equation}\label{state2}
		\begin{aligned}
			\ket{n(t)}&=\cos{\theta} e^{ i \frac{\Delta \omega}{2} t} \ket{n_1} +\sin{\theta} e^{- i \frac{\Delta \omega}{2} t} \ket{n_2}\\
			\ket{n'(t)}&=-\sin{\theta} e^{ i \frac{\Delta \omega}{2} t} \ket{n_1} +\cos{\theta}  e^{- i \frac{\Delta \omega}{2} t} \ket{n_2}
		\end{aligned}
	\end{equation}
	where $\Delta \omega=\omega_2-\omega_1$.
	
	It is known \cite{Geo4,Capolupo2018} that, in presence of mixing, particles spontaneously acquire a \emph{geometric phase} in addition to the dynamical phase generated by their time evolution. We can verify this immediately by computing the neutron phase in the $n-n'$ mixing scenario just discussed. By the kinematic definition \cite{Mukunda} the geometric phase along a given trajectory $\gamma$ is $\Phi^{g}(\gamma) = \Phi^{tot}(\gamma) - \Phi^{dyn}(\gamma)$ where the total phase and the dynamical phase are respectively $\Phi^{tot}(\gamma) = \arg \braket{\psi \left(\lambda_1\right)|\psi \left(\lambda_2\right)}$, $\Phi^{dyn} (\gamma) = \Im{\int_{\lambda_1}^{\lambda_2} { \braket{\psi \left(\lambda\right)|\dot{\psi} \left(\lambda\right)} d\lambda}}$ and $\lambda \in [\lambda_1, \lambda_2]$ is any parametrization of $\gamma$. For a neutron evolving under the action of the mixing Hamiltonian \eqref{H} from $t=0$ we have
	\begin{eqnarray}
                \nonumber && \Phi^{g} \left(t \right) = \arg \braket{n \left(0\right)|n \left(t\right)} - \Im{\int_{0}^{t} { \braket{n \left(t^\prime\right)|\dot{n} \left(t^\prime\right)} dt^\prime}} = \\
                && \arg \left[ \cos\left({\frac{\Delta \omega}{2} t}\right)+i \sin\left({\frac{\Delta \omega}{2} t}\right) \cos{2 \theta } \right] \! - \! \frac{\Delta \omega t}{2} \cos{2 \theta }.
	\end{eqnarray}
	The quantity thus obtained is by definition gauge and reparametrization invariant and it is exclusively related to mixing, vanishing for $\theta = 0$. Since $ \theta = \frac{1}{2}\arctan \left(\frac{2\epsilon_{n n'}}{\delta m-\Delta E}\right)$, it is zero in absence of mixing $\epsilon_{n n'} \rightarrow 0$. Consequently a measurement of $\Phi^{g}$ can be used to test the mirror model, eventually setting bounds on the parameters $\epsilon_{nn'}$ and $\delta m$.
	
	The directly accessible quantities in neutron interferometry are the differences of \emph{total} phases. Yet we can cleverly arrange the interferometric setup in order that the difference of total phases coincides with the difference of geometric phases, by leveraging on the freedom to tune the length of the two arms $l_a, l_b$ and the external magnetic fields. In doing so the observed phase difference and the corresponding interference pattern are only due to the geometric phases, and thus disappear in absence of mixing.

	    \begin{figure}[t]
	\centering
	\includegraphics[width=1\linewidth]{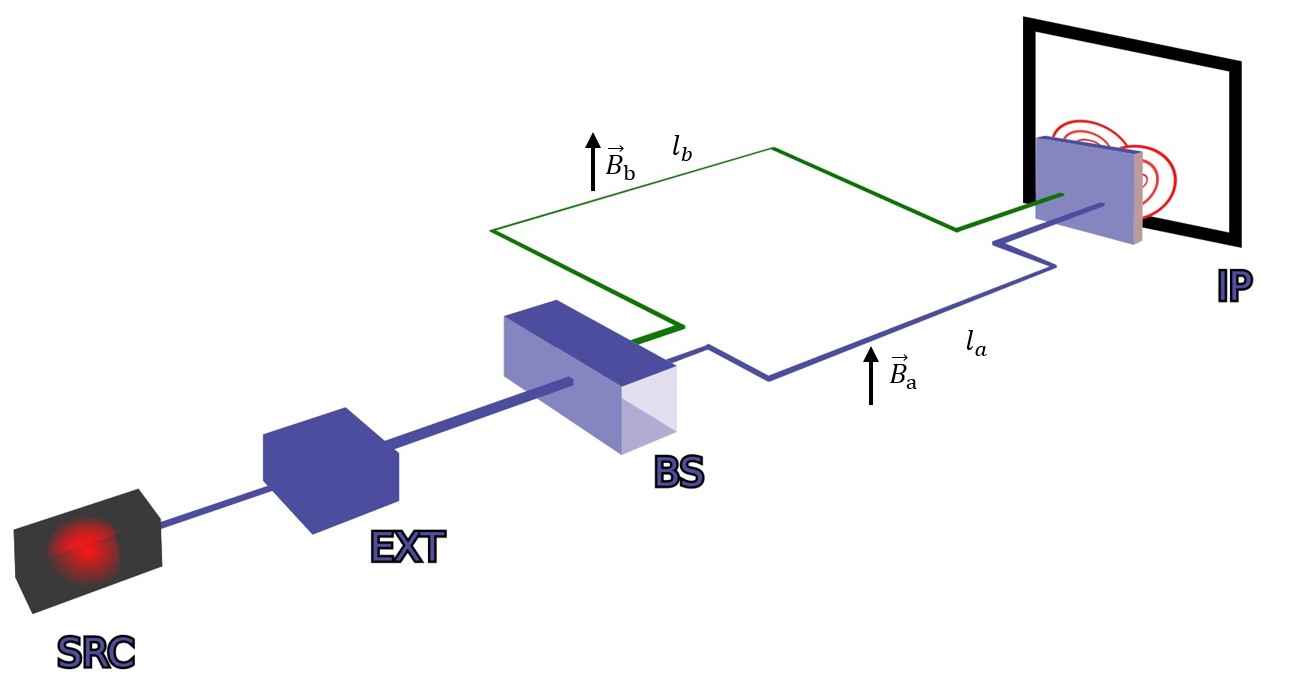}
	\caption{(Color online). Schematic depiction of the interferometric apparatus. The beam is produced by a certain source \textbf{SRC}, polarized through an external device \textbf{EXT}. The beam then goes through a beam splitter \textbf{BS}. Interference is finally observed at the interference plane \textbf{IP}. }
	\label{fig:1}
    \end{figure}

	In particular, we propose the following setup. Denoting by $\hat{y}$ any spatial direction, a neutron beam is polarized in the positive $y$ direction and goes through a beam splitter. The resulting subbeams traverse two arms of different lengths $l_a$ and $l_b$, subject to the action of magnetic fields of different magnitude $B_a$ and $B_b$, concordly oriented along the positive $y$ direction. The two subbeams are then rejoined and interference is observed. A schematic depiction of the setup is shown in Fig. \ref{fig:1}. Provided that the following relation holds among the quantities in the two arms
	\begin{equation}\label{cond}
		l_b=\frac{\Delta \omega_a \cos{2\theta_a}}{\Delta \omega_b \cos{2\theta_b}} l_a
	\end{equation}
	the difference of dynamical phases at the interference plane vanishes, leaving only the difference of geometric phases. Here the angles $\theta_J = \frac{1}{2} \arctan \left(\frac{2\epsilon_{nn'}}{\delta m - |\mu_n B_J|} \right)$ and the energy differences $\Delta \omega_J = \omega_2 (B_J) - \omega_1(B_J)$ depend directly on the magnetic fields for $J=a,b$.

    \begin{figure}[H]
	\centering
	\includegraphics[width=1\linewidth]{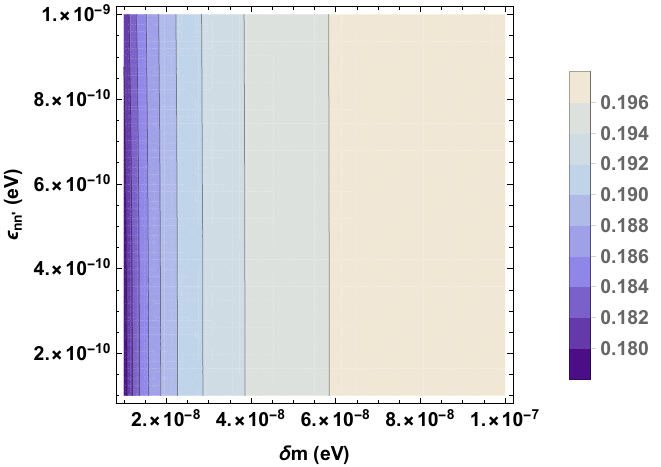}
	\caption{(Color online). Contour plot of $l_b (\mathrm{m})$ from Eq. \eqref{cond} as a function of $\delta m$ and $\epsilon_{n n'}$ for values of the magnetic fields $B_1=2 \cdot 10^{-4} \ \mathrm{T}$ and $B_2=2 \cdot 10^{-2} \ \mathrm{T}$, neutron wavelength $\lambda = 10 \ \mathrm{\AA{}}$ and length of the first arm $l_a = 0.2 \ \mathrm{m}$. The length ratio between the two arms has barely any dependence on $\epsilon_{nn'}$ in the range considered, while it depends strongly on $\delta m$.  }
	\label{fig:2}
    \end{figure}
	
	If the intereferometric setup is arranged such that the condition \eqref{cond} holds, substituting $t = \frac{l_a}{v}$ for the time of flight, where $v$ is the neutron velocity, the observed phase difference is the following difference of geometric phases:
		\begin{equation}\label{diff1}
		\begin{aligned}
			\Delta \Phi &= \Delta \Phi^{g}  =\arg \left[ \cos\left({\frac{\Delta \omega_a}{2 v} \frac{\cos{2\theta_a}}{ \cos{2\theta_b}} l_a}\right)+\right. \\
			& \left. +i \sin\left({\frac{\Delta \omega_a}{2 v} \frac{\cos{2\theta_a}}{ \cos{2\theta_b}} l_a}\right) \cos{2 \theta_2 } \right] - \\
			&-\arg \left[ \cos\left({\frac{\Delta \omega_a}{2 v}l_a}\right)+i \sin\left({\frac{\Delta \omega_a}{2 v}l_a}\right) \cos{2 \theta_a } \right] \ .
		\end{aligned}
	\end{equation}
	It is important to remark once again that such observable bears direct witness to $n-n'$ oscillations, vanishing in absence of mixing.
	
\begin{figure}
	  	\centering
		 \includegraphics[width=1\linewidth]{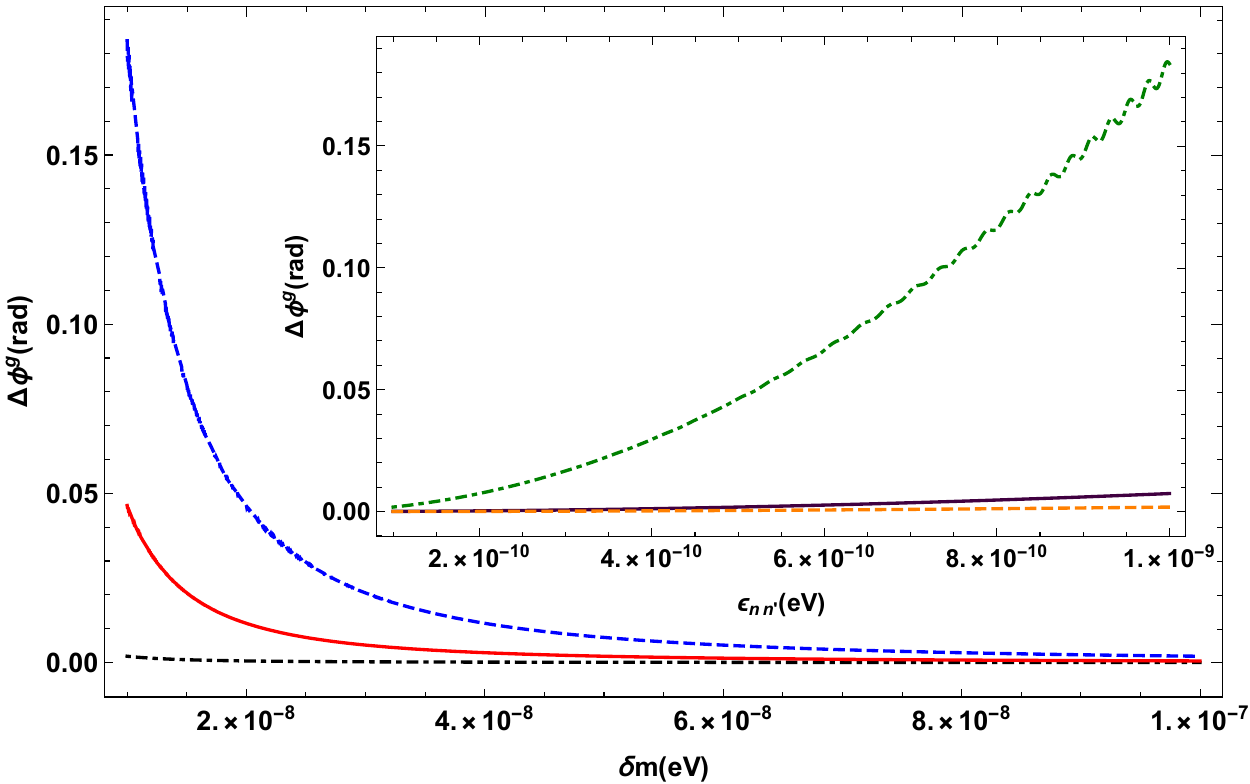}
		 \caption{(Color online). Geometric phase difference $\Delta \phi ^g$ as a function of $\delta m$, for $\epsilon_{n n'}=10^{-10} \ \mathrm{eV}$ (the black dotted line), $\epsilon_{n n'}=5 \cdot 10^{-9} \ \mathrm{eV}$ (the red solid line) and $\epsilon_{n n'}=10^{-9} \ \mathrm{eV}$ (the blue dashed line), and as a function of $\epsilon_{n n'}$ (in the inset) for $\delta m=10^{-8} \ \mathrm{eV}$ (the green dotted line), $\delta m=5 \cdot 10^{-8} \ \mathrm{eV}$ (the solid purple line) and $\delta m=10^{-7} \ eV$ (the orange dashed line). Neutron wavelength $\lambda = 10  \ \mathrm{\AA{}}$, magnetic fields $B_a = 0, B_b = 2 \times 10^{-4} \ \mathrm{T}$ and length of the first arm $l_a = 20 \ \mathrm{cm}$ are considered.}
		\label{fig:3}
		\end{figure}

\begin{figure*}
    	\centering
    	\includegraphics[width=1\textwidth]{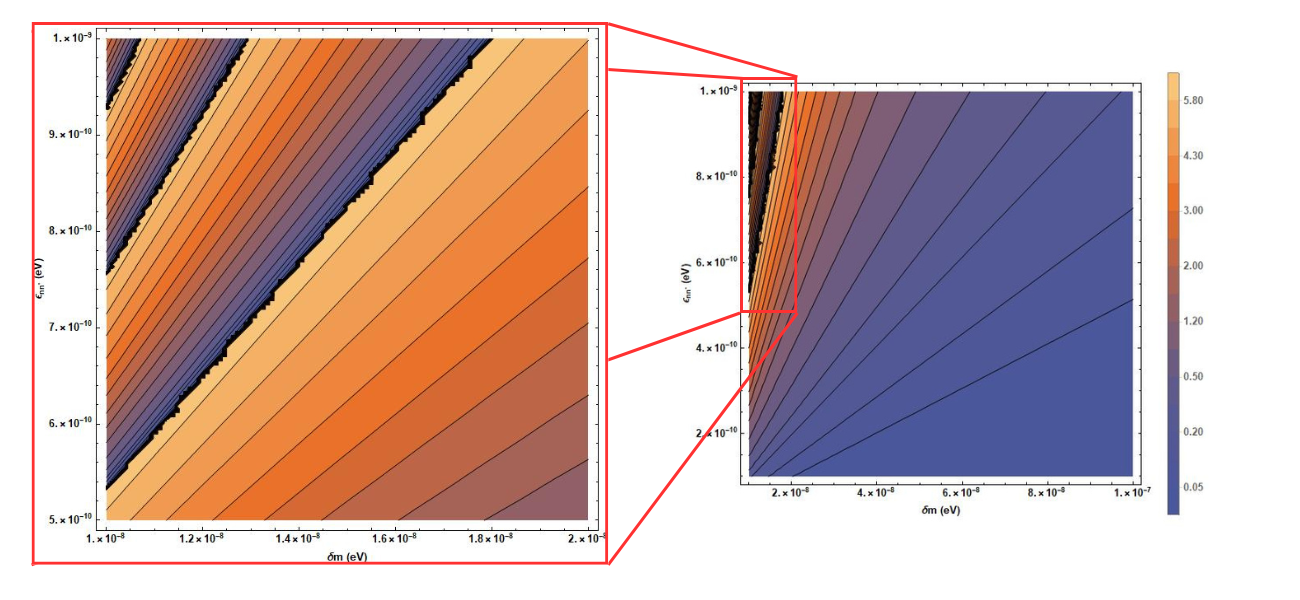}
    	\caption{(Color online). Geometric phase difference $\Delta \phi ^g$ modulo $2\pi$ (in radians) as a function of $\delta m$ and $\epsilon_{n n'}$ for values of the magnetic fields $B_a=2 \times 10^{-4} \ T$ and $B_a=2 \times 10^{-2} \ T$. Wavelength $\lambda = 10  \ \mathrm{\AA{}} $ and arm length $l_a = 20 \ \mathrm{cm}$ are considered.}
    	\label{fig:5}
    \end{figure*}

	To show the behavior of the geometric phase difference \eqref{diff1} we consider parameters compatible with the existing experimental facilities for neutron interferometry. Cold neutron (wavelength $\lambda$ of a few $\mathrm{\AA{}}$) facilities usually involve paths of a few $\mathrm{cm}$ \cite{Facilities}. In figure \ref{fig:3} we plot $\Delta \Phi^{g}$ for arm length $l_a = 20 \ \mathrm{cm}$, wavelength $\lambda =10 \mathrm{\AA{}} $ and magnetic fields $B_a = 0, \ B_b = 2 \times 10^{-4} \mathrm{T}$, as a function of the model parameters $\epsilon_{nn'}$ and $\delta m$.
	It is immediate to see that the geometric phase difference attains larger values for smaller $\delta m$ and larger $\epsilon_{nn'}$. This is immediately understood in terms of the mixing angle $\theta = \frac{1}{2}\arctan \left(\frac{2\epsilon_{n n'}}{\delta m-\Delta E}\right)$, which is indeed maximized for small $\delta m$ and large $\epsilon_{nn'}$. Such behavior is further displayed in the contour plot of Fig. \ref{fig:5} for different values of the magnetic fields $B_a=2 \times 10^{-4} \ \mathrm{T}$ and $B_b=2 \times 10^{-2} \ \mathrm{T}$. The large $\epsilon_{nn'}$-small $\delta m$ region of parameters features several full ($2 \pi$) oscillations of the geometric phase difference. This region is therefore inherently better observable, due to enhancement of the mixing phenomenon.
	\begin{figure}[h]
		\centering
	\includegraphics[width=1\linewidth]{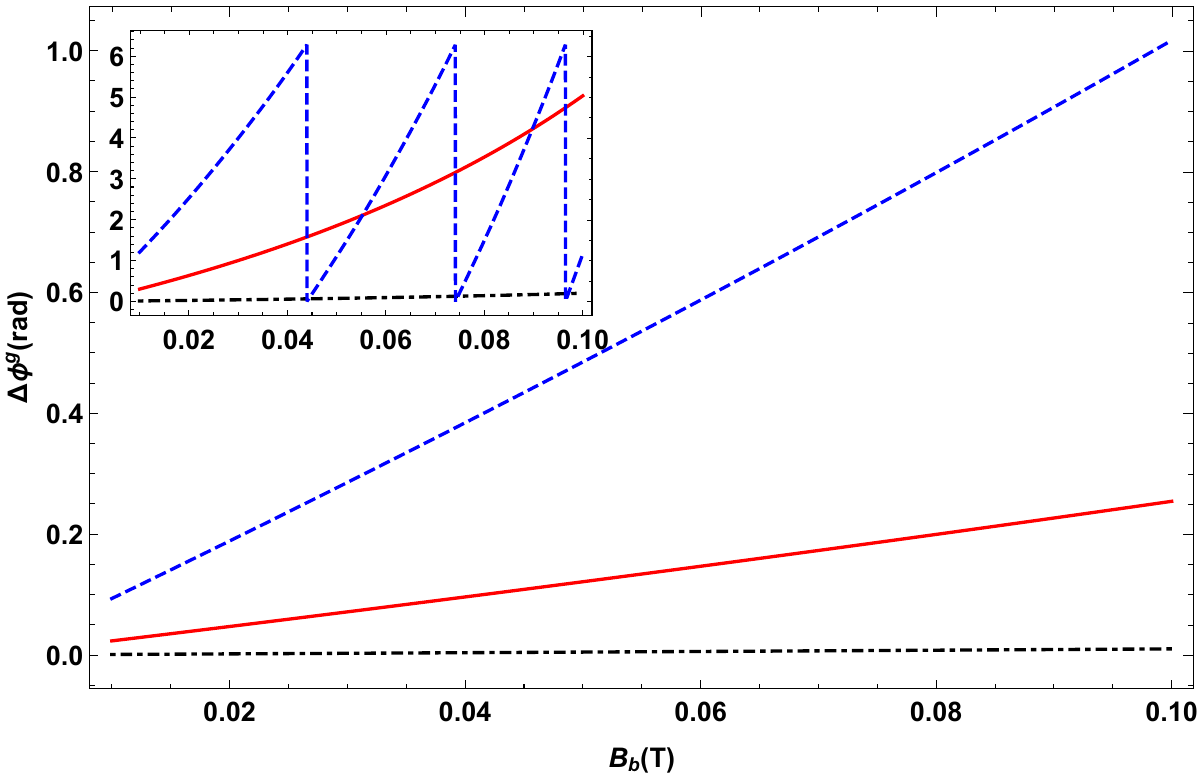}
	\caption{(Color online). Geometric phase difference $\Delta \phi ^g$ modulo $2\pi$ as a function of $B_b$ for $\epsilon_{n n'}=10^{-10} \ \mathrm{eV}$ (the black dotted line), $\epsilon_{n n'}=5 \cdot 10^{-9} \ \mathrm{eV}$ (the red solid line) and $\epsilon_{n n'}=10^{-9} \ \mathrm{eV}$ (the dashed blue line); we fix $\delta m=10^{-7} \ \mathrm{eV}$( $\delta m=2 \times 10^{-8} \ \mathrm{eV}$ in the inset), $l_a = 20 \ \mathrm{cm}$, $\lambda = 10 \ \mathrm{\AA{}}$ and $B_a = 0$.}
	\label{fig:4}
    \end{figure}
	A similar argument holds also for the magnetic fields. As the energy shift $|\mu_{n} B|$ approaches the resonance threshold $\simeq \delta m$, mixing is significantly enhanced. In practice the resonance condition is met around $B \sim 1 \ \mathrm{T}$ for $\delta m \simeq 10^{-7} \ \mathrm{eV}$. As a consequence, the geometric phase difference can grow large enough to complete several full oscillations as the magnetic field is increased. This is shown in Fig. \ref{fig:4}.

Arm length can be likewise increased to enhance the geometric phase difference, in particular for the region of parameters (small $\epsilon_{nn'}$-large $\delta m$), for which mixing is suppressed. The usage of tinier magnetic fields and shorter lengths is instead desirable in the region of parameters for which mixing is enhanced, to avoid a strong oscillatory behavior of the geometric phase. Variable geometries and magnetic fields can be implemented to better probe different regions of the parameter space.
Limitations to the observability of the phase difference of Eq. \eqref{diff1} are posed by additional unwanted sources of phase shifts. The main two sources in this case are represented by curvature and spin precession. A certain curvature within at least one of the two arms is required to let the subbeams join at the interference plane, because $l_a \neq l_b$ is forced by the condition \eqref{cond}. While this unavoidably adds an unwanted phase shift to at least one of the two paths, it is likewise the case that it can easily be controlled and properly subtracted to obtain the result of Eq. \eqref{diff1}. A simple way to do so is to observe the interference pattern in absence of magnetic fields, or for $B_a = B_b$, in which case \eqref{diff1} vanishes and the residual phase shift is exclusively due to the geometry of setup. Other minor sources of noise, such as the effect of mirrors and spurious interactions with matter can be kept under control in the same way.
    If the neutron beam is not perfectly polarized with respect to the common direction of the magnetic fields $\hat{y}$, and a mixture of the two polarizations enters the area with magnetic fields, an additional phase shift due to spin precession is generated. Since the intensities are required to be different $B_a \neq B_b$, an extra phase difference would be observed with respect to \eqref{diff1}. However also this second limitation can be easily kept under control, simply by tuning the arm lengths, depending on the spin composition of the beam, so that also the magnetic phase shift is removed.

The proposed setup may probe, with minimal limitations, a significant portion of parameter space in neutron-mirror neutron oscillations, observing the geometric phase that naturally accompanies the phenomenon.
    The discovery of such a phase could open up unexplored scenarios, revealing a yet unobserved sector of particles and interactions, possibly contributing to the dark sector of the universe and involved in experimental anomalies in particle physics.

\section*{Acknowledgements}
We acknowledge partial financial support from MUR and INFN, A.C. also acknowledges the COST Action CA1511 Cosmology
and Astrophysics Network for Theoretical Advances and Training
Actions (CANTATA).

\end{document}